\documentclass[preprint, showpacs, preprintnumbers, amsmath, amssymb]{revtex4}
\usepackage{graphicx}
\usepackage{dcolumn}
\usepackage{bm}
\usepackage{epsfig}

\begin{document}

\title{SELF-DUAL VORTICES IN THE FRACTIONAL QUANTUM HALL SYSTEM}
\author{XIN-HUI ZHANG}
\thanks{Corresponding author}%
\email{zhangxingh03@lzu.cn}%
\author{YI-SHI DUAN}
\author{YU-XIAO LIU}
\author{LI ZHAO}
\affiliation{ Institute of Theoretical Physics, Lanzhou University,
Lanzhou 730000, P. R. China}

\begin{abstract}

Based on the $\phi$-mapping theory, we obtain an exact Bogomol'nyi
self-dual equation with a topological term, which is ignored in
traditional self-dual equation, in the fractional quantum Hall
system. It is revealed that there exist self-dual vortices in the
system. We investigate the inner topological structure of the
self-dual vortices and show that the topological charges of the
vortices are quantized by Hopf indices and Brouwer degrees.
Furthermore, we study the branch processes in detail. The vortices
are found generating or annihilating at the limit points and
encountering, splitting or merging at the bifurcation points of the
vector field $\vec\phi$.
\end{abstract}


\maketitle
%74.20.De Phenomenological theories (two-fluid, Ginzburg-Landau, etc.)
%73.43.-f Quantum Hall effects

\section{Introduction}

The fractional quantum Hall (FQH) effect is an example of the new
physics that has emerged from the enormous progress made during the
past few decades in material synthesis and device processing
\cite{H.L.StormerRMP1999,K.V.KlitzingPRL1980}. Since its discovery
\cite{D.C.TsuiPRL1982}, experiments on FQH systems have continued to
reveal many new phenomena and surprises. Filling factors may take
the values $\nu={1}/(2m\pm{1})$ or $\nu=m/(2mp\pm{1})$, in which $m$
and $p$ are integers. All these spark interest to theoretical work
of the FQH system. In 1983 Laughlin proposed his celebrated wave
functions as an explanation of the FQH effect for two-dimensional
electron gas with filling factors $\nu=1/(2m\pm{1})$
\cite{R.B.LaughlinPRL1983}. Shortly after that, hierarchical
generalizations of these fully polarized states were also proposed
for filling factors $m/(2mp\pm{1})$, with odd-denominator fractions
\cite{B.I.HalperinPRL1984,F.D.M.HaldanePRL1983}. Subsequently, Jain
pointed out that the FQH effect of electrons can be physically
understood as a manifestation of the integer quantum Hall effect of
composite fermionic objects consisting of electrons bound to an even
number of flux quanta \cite{J.K.Jain}. However, the observation of
fractions such as $4/11$ and $5/13$
\cite{W.PanIJMPB2002,J.H.SmetNature2003} points to new physics
beyond the integral quantum Hall effect of composite fermions.
Recently, Ref. \cite{W.panPRL2003} presents extensive experimental
evidence for the considerable strength of these interactions
observing the appearance of FQH effect states at filling factors
$\nu=7/11$, $4/13$, $6/17$ and $15/17$, located between the minima
of the primary FQH effect sequences.

Despite the success of the microscopic theories, it is important to
develop an effective-field-theory model analogous to the
Ginzburg-Landau theory of superconductivity. Zhang, Hansson and
Kivelson constructed the bosonic Chern-Simon field theory (ZHK
model) of the Laughlin states describing the FQH states as the Bose
condensation of a (bosonic) field \cite{AnaPRB2004}, from which all
the essential features can be derived \cite{S.C.ZhangPRL1989}.
Girvin and MacDonald \cite{M.GirvinPRL1987} and Read
\cite{N.ReadPRL1989} also discovered a hidden order parameter in the
FQH system. They proposed a field-theory model, containing a complex
scalar field $\phi$ coupled to a vector field $(a_0,\vec{a})$ with a
Chern-Simon action (or topological mass term). This model exhibits
vortex solutions with finite energy and fractional charge which can
be identified with Laughlin's quasiparticles and quasiholes. By
adding a natural magnetic term to the field-theory model of Girvin
and MacDonald, Ref. \cite{M.HassaineJPA1998} has shown that the FQH
system admits stable topological as well as non-topological vortex
solutions and the physical significance of the ``topological"
vortices corresponds to quasiparticles and quasiholes. These
successes certainly make the effective-field approach extremely
appealing. Recently, the intrinsic spin Hall effect has been
theoretically predicted for semiconductors with spin-orbit coupled
band structures \cite{S.MurakamiScience2003,J.SinovaPRL2004}. It has
been argued that the spin quantum Hall liquid is a novel state of
matter with profound correlated properties described by a
topological field theory \cite{B.A.BernevigPRL2006}.

We can see that the topological properties play an important role in
the FQH system. The purpose of the paper is to study the FQH system
from the point of view of topology. We introduce a topological
method, the $\phi$-mapping topological current theory, which
provides a powerful method in researching some topological
properties. It has been effectively used to study topological
characteristic of dislocations and disclinations continuum
\cite{Y.S.DuanIJES1990}, the topological properties of magnetic
monopoles in superconductors \cite{Y.JiangPRB2004} and the knotted
soliton in two-gap superconductors \cite{ZhangPRB}. By making use of
the $\phi$-mapping theory, the self-dual vortices are investigated
in the frame of the ZHK model in the FQH system. The paper is
arranged as follows. In Sec. \ref{section1}, by studying the
Bogomol'nyi first-order self-dual equation, we point out that there
exist self-dual vortices in the FQH system and research their inner
topological structure. In Sec. \ref{section2}, we study the
evolution of the self-dual vortices. The conclusion of this paper is
given in Sec. \ref{conclusion}.

\section{self-dual vortices in the FQH system}\label{section1}

The FQH effect appears in a two-dimensional electron system in a
strong magnetic field. It is known that in (2+1)-dimensional
spacetime the kinetic action for a gauge field can be either the
Maxwell term or the Chern-Simon term, or both. Ref.
\cite{E.B.Bogomol'nyiYF1976SJNP1976} has shown that there can be a
Bogomol'nyi-type bound  for the energy functional in a pure Abelian
Chern-Simon theory. In this paper, we start with the Lagrangian of
the ZHK model \cite{S.C.ZhangPRL1989}
\begin{eqnarray}
L_{ZHK}&=&-\frac{\kappa}{2}\epsilon^{\mu\nu\rho}a_\mu
\partial_\nu{a_\rho}+i\phi^*(\partial_0+ia_0)\phi \nonumber\\
&-&\frac{1}{2m}|(\partial_i+i(a_i+A_i^{ext}))\phi|^2\\
&-&\frac{1}{2}\int{d}^2x'(|\phi(\vec{x})|^2-n)V(\vec{x}-\vec{x'})
(|\phi(\vec{x'})|^2-n)\nonumber,
\end{eqnarray}
which just contains the pure Abelian Chern-Simon term. Here $a_\mu$
is the statistical Chern-Simon gauge field, the external gauge field
$A^{ext}_i$ describes the external magnetic field, the constant $n$
denotes a uniform condensate charge density. Since in a
two-dimensional system, a spinless electron may be represented as a
hardcore boson carrying an odd integer of Dirac flux quanta, when
the Chern-Simon coupling $\kappa$ takes the values
\begin{equation}\label{k}
\kappa=\frac{1}{2\pi(2N-1)}\;\;\;(N\geq{1}),
\end{equation}
one can regard this as the condensing of the fundamental fermion
field into bosons by the attachment of an odd number of flux through
the Chern-Simon coupling \cite{S.C.ZhangPRL1989}. We consider a
$\delta$-function contacted interaction with
$V(\vec{x}-\vec{x'})=\frac{1}{mk}\delta(\vec{x}-\vec{x'})$, then the
potential can be written as $V(\rho)=\frac{1}{2mk}(\rho-n)^2$, in
which $\rho=\phi^*\phi$. The static energy functional for this model
is
\begin{eqnarray}
\varepsilon_{ZHK}&=&\int{d}^2x \left[\frac{1}{2m}|(\partial_i
+i(a_i+A^{ext}_i))\phi|^2\right.\\\nonumber
&&+\left.\frac{1}{2mk}(\rho-n)^2\right].
\end{eqnarray}
Clearly, the minimum energy solutions correspond to the constant
field solutions $\phi=\sqrt{n},\;a_i=-A_i^{ext},\;a_0=0$. In this
case the Chern-Simon gauge field opposes and cancels the external
field, thus the FQH effect can be viewed as the condensation of the
hardcore bosons. Since the Chern-Simon constraint is
$b=-\frac{1}{\kappa}\rho$ \cite{G.V.Dunne1998} (where $b$ is the
Chern-Simon gauge field tensor), we learn that the minimum energy
solutions have density $\rho=n=\kappa{B^{ext}}$. Here $\kappa$ takes
the values of Eq. (\ref{k}). These are exactly the conditions for
the uniform Laughlin states of filling fraction
$\nu=\frac{1}{2N-1}$, at which there exist $(2N-1)$ times as many
vortices as there are electrons, each vortex representing a local
charge deficit $\frac{e}{2N-1}$. According to the identity
$|\vec{D}\phi|^2=|(D_1\pm{iD}_2)\phi|^2\mp{eB}|\phi|
^2\pm\epsilon^{ij}\partial_iJ_j$, in which
$J_j=\frac{1}{2i}[\phi^*D_j\phi-\phi(D_j\phi)^*]$, the static energy
can be reexpressed as
\begin{eqnarray}
\varepsilon_{ZHK}&=&\int{d}^2x
\left[\frac{1}{2m}|D_{\pm}\phi|\nonumber
^2\mp\frac{1}{2m}(B^{ext}-\frac{1}{\kappa}\rho)\rho\right.\\
&&+\left.\frac{1}{2mk}(\rho-n)^2 \right]\\ \nonumber  &=&\int{d^2x}
\left[\frac{1}{2m}|D_{\pm}\phi|^2\pm\frac{1}{2mk}
(\rho-\kappa{B}^{ext})^2\right. \\ \nonumber
&&\mp\left.\frac{\kappa}{2m}B^{ext}B + \frac{1} {2mk}(\rho-n)^2
\right]\\ \nonumber
&=&\int{d^2x}\left[\frac{1}{2m}|D_{\pm}\phi|^2\mp\frac{\kappa}
{2m}B^{ext}B \right], \nonumber
\end{eqnarray}
where the total magnetic $B=B^{ext}+b$ and
$D_\pm\phi=(D_1\pm{D_2})\phi$. Then the energy is bounded below by a
multiple of the total magnetic flux and obeys a Bogomol'nyi-type
lower bound, which is achieved by the field satisfying a set of
first-order self-dual equations
\begin{eqnarray}\label{selfdual}
 D_{\pm}\phi=0, ~~~B=B^{ext}-\frac{1}{\kappa}\rho.
\end{eqnarray}

Furthermore, by investigating the above Bogomol'nyi self-dual
equations, we will see there exist the self-dual vortices in the FQH
system. It is known that the complex scalar field
$\phi=\phi^1+i\phi^2$ can be regard as the complex representation of
a two-dimensional vector field $\vec{\phi}=(\phi^1,\phi^2)$ over the
base manifold, and it is actually a section of a complex line bundle
on the base spacetime manifold. From the two-dimensional vector
field $\vec{\phi}=(\phi^1,\phi^2)$, we can define the unit vector
field
\begin{equation}
n^a=\frac{\phi^a}{\|\phi\|},\;\;\;\|\phi\|^2=\phi^*\phi.
\end{equation}
This is a reasonable representation, $\phi^a$ is a two component
vector field related to the order parameter field $\phi$. Obviously,
it can be looked upon as a smooth mapping between the
two-dimensional space $X$ (with the local coordinate $x$) and the
two-dimensional Euclidean space $R^2$, $\phi$ :
$x\mapsto\vec\phi(x)\in{R^2}$ and $n^a$ a section of the sphere
bundle $S(x)$. Clearly, the zero points of the $\vec{\phi}$ field
are just the singular points of the unit vector field. In the
following, by virtue of the so-called $\phi$-mapping theory, we will
point out that, when $\phi^a$ field possesses several zero points,
there exists a topological term, taking the form of the
$\delta$-function. From the first case of self-dual equations
$D_{+}\phi=0$, by separating the real part from the imaginary, we
obtain $eA_\mu=\epsilon_{ab}n^a\partial_\mu{n^b}-({1}/{2})\epsilon
^{\mu\nu}\partial_\nu\ln(\phi^*\phi)$. From  the second case, i.e.
$D_{-}\phi=0$, by repeating the same process, we get
$eA_\mu=\epsilon_{ab}n^a\partial_\mu{n}^b+({1}/{2})\epsilon
^{\mu\nu}\partial_\nu\ln(\phi^*\phi)$. Then we have
$eA_\mu=\epsilon_{ab}n^a\partial_\mu{n}^b\pm\frac{1}{2}
\epsilon^{\mu\nu}\partial_\nu\ln(\phi^*\phi)$. In (2+1)-dimensional
spacetime, the magnetic field is $B=\epsilon^{ij}\partial_iA_j$.
According to above equations, we have
\begin{equation}
B=\epsilon^{\mu\nu}\epsilon_{ab}\partial_\mu{n^a}\partial_\nu{n^b}
+\nabla^2\ln{\|\phi\|^2}.
\end{equation}
Noticing the relation
$\partial_\mu{n^a}=({\partial_\mu\phi^a})/{\|\phi\|}
+\phi^a\partial_\mu({1}/{\|\phi\|})$ and the well-known Green
function relation in $\phi$-space
$\partial_a\partial_a\ln\|\phi\|=2\pi \delta^2(\vec{\phi})\
(\partial_a={\partial}/{\partial\phi^a})$, one can  prove that
$\epsilon^{\mu\nu}\epsilon_{ab}\partial_\mu{n^a}\partial_\nu{n^b}
=2\pi\delta^2(\vec\phi)D({\phi}/{x})$, where
$D(\phi/x)=\frac{1}{2}\epsilon^{\mu\nu}\epsilon_{mn}
(\partial\phi^m/\partial{x^\mu})(\partial\phi^n/\partial{x^\nu})$ is
the Jacobian. So the Bogomol'nyi self-dual equation (\ref{selfdual})
can be written as
\begin{equation}\label{self1}
2\pi\delta^2(\vec{\phi})D(\frac{\phi}{x})+\nabla^2\ln\rho=
B^{ext}-\frac{1}{\kappa}\rho,
\end{equation}
where $\rho=\|\phi\|^2$ is the field density. This equation is more
exact than the usual self-dual equation in the ZHK model
\cite{G.V.Dunne1998}
\begin{equation}\label{usualself}
\nabla\ln^2\rho=B^{ext}-\frac{1}{\kappa}\rho,
\end{equation}
in which the first term $2\pi\delta^2(\vec{\phi})D(\frac{\phi}{x})$
on the LHS of Eq. (\ref{self1}) has been ignored all the time. From
our previous work, obviously, the first term of Eq. (\ref{self1})
describes the topological properties of the self-dual FQH system. As
for usual self-dual equation (\ref{usualself}), it only describes
the non-topological properties of the system.

In order to investigate the topological properties of the FQH system
in more detail, let's introduce a topological current in terms of
the topological term
\begin{equation}\label{current}
J^\mu=\frac{1}{2\pi}\epsilon^{\mu\nu\lambda}\epsilon_{ab}
\partial_\nu{n^a}\partial_\lambda{n^b}
=\delta^2(\vec{\phi})D^\mu(\frac{\phi}{x}),
\end{equation}
where
$D^\mu(\phi/x)=({1}/{2})\epsilon^{\mu\nu\lambda}\epsilon_{mn}
(\partial\phi^m/\partial{x^\nu})(\partial\phi^n/\partial{x^\lambda})$
is the Jacobian vector. From Eq. (\ref{current}), we can  see that
the topological current does not vanish only at the zero points of
the $\vec\phi$ field. So it is necessary to study the zero points
of the $\vec\phi$ field to determine the nonzero solution of the
topological current. The implicit function theory \cite{implicit}
shows that under the regular condition $D^\mu({\phi}/{x})\neq{0}$,
the general solutions of
\begin{equation}\label{phi}
\phi^1(x^1,x^2,t)=0,\ \ \ \phi^2(x^1,x^2,t)=0,
\end{equation}
can be expressed as $x^a=x^a_k(t)\ (k=1,2,\cdots,N)$, which
represent the world lines of $N$ moving isolated singular points.
These singular  solutions are just the self-dual vortices in the FQH
system. In $\delta$-function theory \cite{delta}, one can prove that
\begin{equation}\label{delta}
\delta^2(\vec\phi)=\sum^{N}_{k=1}\frac{\beta_k}{|D(\frac{\phi}{x})
|_{\vec{x}_k}} \delta^2(\vec{x}-\vec{x}_k).
\end{equation}
Here the positive integer $\beta_k$ is the Hopf index of the
$\phi$-mapping, which means that when $\vec{x}$ covers the
neighborhood of the zero point $\vec{x}_k$ once, the vector field
$\vec\phi$ covers the corresponding region in $\phi$ space $\beta_k$
times. With the definition of vector Jacobian, we can obtain the
general velocity of the $k$-th vortices
\begin{equation}\label{v}
v^\mu_k=\left.\frac{dx^\mu_k}{dt}=\frac{D^\mu(\phi/x)}
{D(\phi/x)}\right|_{\vec{x}_k},\ \ v^0=1.
\end{equation}
Then the topological current $J^\mu$ can be written as the form of
the current and the density of the system of $N$ classical point
particles with topological charge $W_l=\beta_l\eta_l$ moving in the
(2+1)-dimensional spacetime
\begin{eqnarray}\label{current2}
\vec{J}&=&\sum^N_{k=1}\beta_k\eta_k\vec{v}_k\delta^2(\vec{x}-\vec{x}_k),\nonumber\\
\rho&=&J^0=\sum^N_{k=1}\beta_k\eta_k\delta^2(\vec{x}-\vec{x}_k),
\end{eqnarray}
where $\eta_k=\textrm{sgn}(D(\phi/x)|_{\vec{x}_k})=\pm{1}$ is the
Brouwer degree of the $\phi$-mapping. It is clear to see that Eq.
(\ref{current2}) shows the movement of the self-dual vortices in
spacetime. So, the total charge of the FQH system can be written as
\begin{equation}
Q=\int\rho(x)d^2x=\sum_{k=1}^{N}\beta_k\eta_k.
\end{equation}
It is obvious to learn that there exist $N$ isolated vortices, of
which the $k$-th vortex possesses charge $\beta_k\eta_k$. And
$\eta_k=+1$ corresponds to the vortex, while $\eta_k=-1$ corresponds
to the antivortex.

\section{the evolution of the self-dual vortices}\label{section2}

\begin{figure}[h]
\begin{center}
\begin{tabular}{p{5cm}p{5cm}}
\psfig{figure=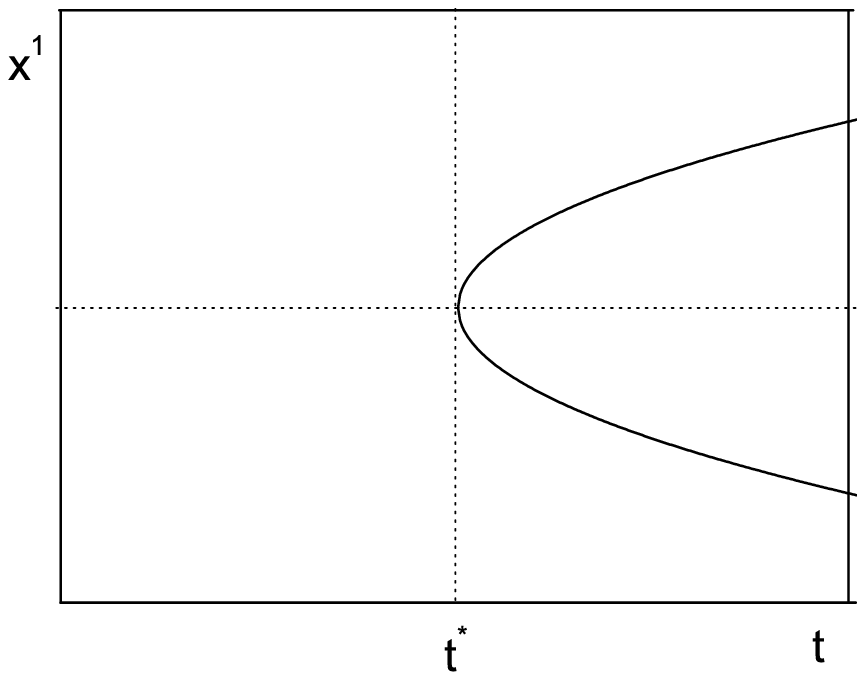,height=5cm,width=5cm} &
\psfig{figure=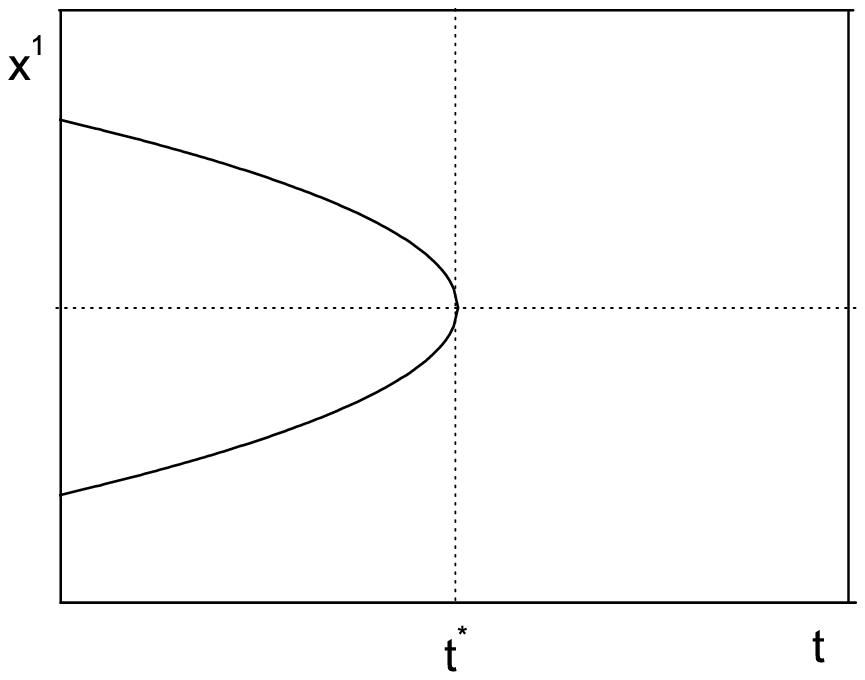,height=5cm,width=5cm}\\
\hspace{2.8cm} (a) & \hspace{2.8cm} (b)
\end{tabular}
\end{center}
\caption{ (a) A pair of vortices with opposite charges generate at
the limit point, i.e., the origin of vortices. (b) A pair of
vortices with opposite charges annihilate at the limit point.}
\label{fig1}
\end{figure}

In the above section, we have studied the topological properties
of the self-dual vortices in the case that the vector order
parameter $\vec\phi$ only consists of regular points, i.e.,
$D^{\mu}({\phi}/{x})\neq0$ is hold true. However, when the regular
condition fails, branch processes will occur. Usually there are
two kinds of branch points, namely the limit points and the
bifurcation points. In this section, we will study the evolution
of the self-dual vortices in the FQH system. From Eq. ({\ref{v}}),
we can learn that the velocity of the $k$-th zero point is
determined by
\begin{equation}\label{velocity2}
\frac{dx^{i}}{dt}=\left.\frac{D^{i}(\frac{\phi}{x})}{D^0
(\frac{\phi}{x})} \right|_{x=\vec{x}_{k}},\;\;(i=1,2),
\end{equation}
where $D^0(\phi/x)=D(\phi/x)$ is the usual Jacobian.  It is
obvious that when $D^{0}({\phi}/{x})=0$, at the very point
$({t^{*}, \vec{x}^{*}})$ the velocity
%\begin{eqnarray}
%\frac{dx^{1}}{dt}=\left.\frac{D^{1}(\frac{\phi}{x})}
%{D^{0}(\frac{\phi}{x})}\right|_{(t^{*}, \vec{x}^{*})},\ \ \
%\frac{dx^{2}}{dt}=\left.\frac{D^{2}(\frac{\phi}{x})}
%{D^{0}(\frac{\phi}{x})}\right|_{(t^{*}, \vec{x}^{*})},
%\end{eqnarray}
is not unique in the neighborhood of $(t^{*}, \vec{x}^{*})$. If the
Jacobian
$\left.D^1(\frac{\phi}{x})\right|_{(t^*,\vec{x}^*)}\neq{0}$, we can
use the Jacobian $D^1(\phi/x)$ instead of $D^0(\phi/x)$ for the
purpose of using the implicit function theorem \cite{implicit}. Then
we have a unique solution of Eqs. (\ref{phi}) in the neighborhood of
the very point $(t^*,\vec{x}^*)$
\begin{equation}
t=t(x^1),\ \ x^2=x^2(x^1).
\end{equation}
We call the critical points $(t^*,\vec{x}^*)$ the limit points. In
the present case, we know that
\begin{equation}
\frac{dx^{1}}{dt}=\left.\frac{D^{1}(\frac{\phi}{x})}
{D^{0}(\frac{\phi}{x})}\right|_{(t^{*}, \vec{x}^{*})}=\infty,\ \ \
\left.\frac{dt}{dx^1}\right|_{(t^*,\vec{x}^*)}=0.
\end{equation}
Then, the Taylor expansion of $t=t(x^1)$ at the limit point
$(t^*,\vec{x}^*)$ is
\begin{equation}\label{talor}
t-t^*=\frac{1}{2}\left.\frac{d^2t}{(dx^1)^2}\right|_{(t^*,\vec{x}^*)}
(x^1-x^{1*})^2,
\end{equation}
which is a parabola in $x^1-t$ plane. From Eq. (\ref{talor}) we can
obtain two solutions $x^1_1$ and $x^1_2$, which give two branch
solutions (world lines of vortices). If $\frac{d^2t}{(dx^1)^2}>0$,
we have the branch solutions for $t>t^*$; otherwise, we have the
branch solutions for $t<t^*$ [see FIG. \ref{fig1}]. These two cases
are related to the generation and annihilation of the self-dual
vortices.

Since the topological current is identically conserved, the
topological charges of these two generated or annihilated vortices
must be opposite at the limit point, i.e.,
\begin{equation}
\beta_{k_{1}}\eta_{k_{1}}=-\beta_{k_{2}}\eta_{k_{2}},
\end{equation}
which shows that $\beta_{k_{1}}=\beta_{k_{2}}$ and
$\eta_{k_{1}}=-\eta_{k_{2}}$. One can see that the Brouwer degree
$\eta_k$ is indefinite at the limit points, i.e. it can change
discontinuously at limit points along the world lines of self-dual
vortices (from $\pm{1}$ to $\mp{1}$).

In the following, the more complicated case will be discussed. We
have the restrictions of  Eq. (\ref{phi}) at the bifurcation point
$(t^*,\vec{x}^*)$,
\begin{equation}
\left.D^1\left(\frac{\phi}{x}\right)\right|_{\vec{x}^*}=0,\ \ \
\left.D^2\left(\frac{\phi}{x}\right)\right|_{\vec{x}^*}=0.
\end{equation}
Without loss of generality, we discuss only the branch of the
velocity component $(dx^{1}/dt)$ at $(t^{*}, \vec{x}^{*})$. It is
known that the Taylor expansion of the solutions of Eq. (\ref{phi})
in the neighborhood of $(t^{*}, \vec{x}^{*})$ can generally be
expressed as
\begin{equation}
A(x^{1}-x^{1*})^{2}+2B(x^{1}-x^{1*})(t-t^{*})+C(t-t^{*})^{2}+\cdots=0,
\end{equation}
where $A, B$ and $C$ are three constants. Then the above Taylor
expansion can lead to
\begin{equation}\label{Taylor}
A\left(\frac{dx^{1}}{dt}\right)^{2}+2B\left(\frac{dx^{1}}
{dt}\right)+C=0,
\end{equation}
and
\begin{equation}\label{Taylor2}
C\left(\frac{dt}{dx^1}\right)^2+2B\frac{dt}{dx^1}+A=0.
\end{equation}
The solutions of Eqs. ({\ref{Taylor}}) and (\ref{Taylor2}) give
different motion directions of the zero point at the bifurcation
point. There are four possible cases, which will show the physical
meanings of the bifurcation points.

\begin{figure}[h]
\begin{center}
\begin{tabular}{p{5cm}p{5cm}p{5cm}}
\psfig{figure=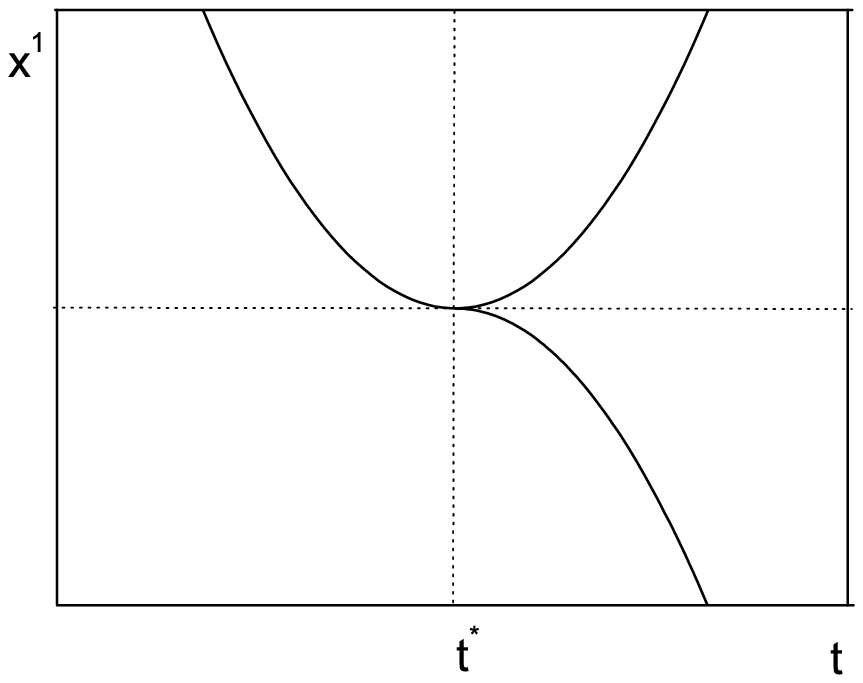,height=5cm,width=5cm} &
\psfig{figure=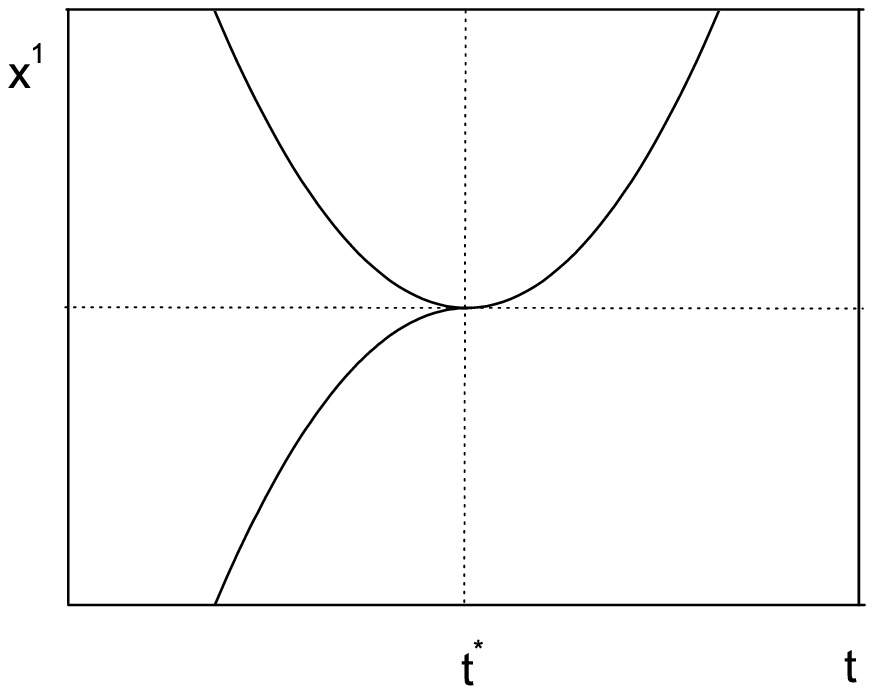,height=5cm,width=5cm} &
\psfig{figure=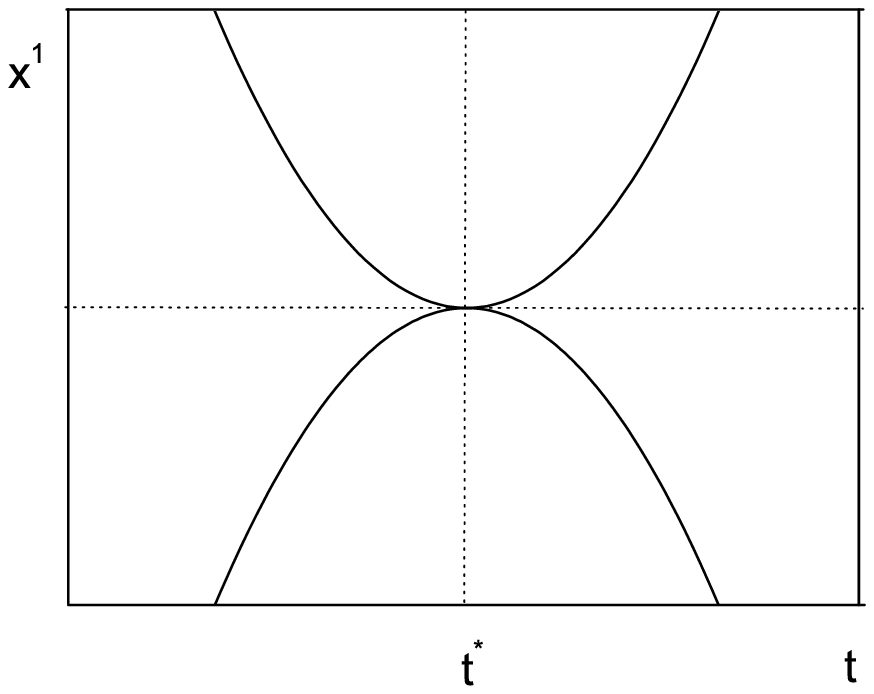,height=5cm,width=5cm} \\
 \hspace{2.3cm} (a)
& \hspace{2.3cm} (b) & \hspace{2.3cm} (c)
\end{tabular}
\end{center}
 \caption{ (a) One vortex splits into two vortices. (b) Two
vortices merge into one. (c) Two vortices tangentially intersect.}
\label{fig2}
\end{figure}

Case 1 ($A\neq{0}$). For $\bigtriangleup=4(B^{2}-AC)=0$, from Eq.
(\ref{Taylor}), we get only one motion direction of the zero point
at the bifurcation point: $(dx^{1}/dt)|_{1, 2}=-B/A$, which includes
three sub-cases: (a) one vortex splits into two vortices; (b) two
vortices merge into one; (c) two vortices tangentially intersect at
the bifurcation point [see FIG. \ref{fig2}].

Case 2 ($A\neq{0}$). For $\bigtriangleup=4(B^{2}-AC)>0$, from Eq.
(\ref{Taylor}), we get two different motion directions of the zero
point: $(dx^{1}/dt)=(-B\pm\sqrt{B^{2}-AC})/A$, which is shown in
FIG. \ref{fig2}. This is the intersection of two vortices, which
means that the two vortices  meet and then depart at the bifurcation
point.

\begin{figure}[h]
\begin{center}
\begin{tabular}{p{5cm}p{5cm}}
\psfig{figure=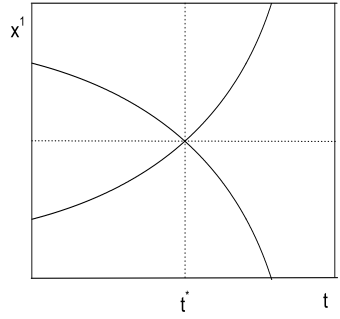,height=5cm,width=5cm}& \hspace{2.8cm}
\end{tabular}
\end{center}
\caption{Two vortices meet and then depart at the bifurcation
point.}\label{fig3}
\end{figure}

Case 3 ($A=0,C\neq{0}$). For $\bigtriangleup=4(B^{2}-AC)>0$, from
Eq. (\ref{Taylor2}) we have
\begin{equation}
\left.\frac{dt}{dx^1}\right|_{1,2}=\frac{-B\pm\sqrt{B^2-AC}}{C}=0, \
\ -\frac{2B}{C}.
\end{equation}
There are two important cases: (a) one world line resolves into
three world lines, i.e., one vortex splits into three vortices at
the bifurcation point. (b) Three world lines merge into one world
line, i.e., three vortices merge into one vortex at the bifurcation
point [see FIG. \ref{fig3}].

\begin{figure}[h]
\begin{center}
\begin{tabular}{p{5cm}p{5cm}}
\psfig{figure=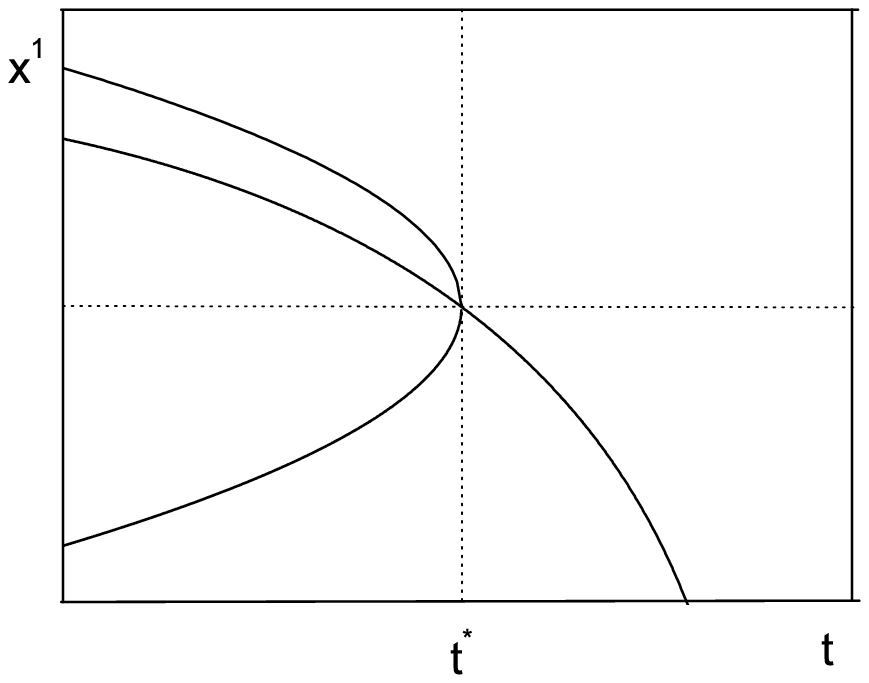,height=5cm,width=5cm} &
\psfig{figure=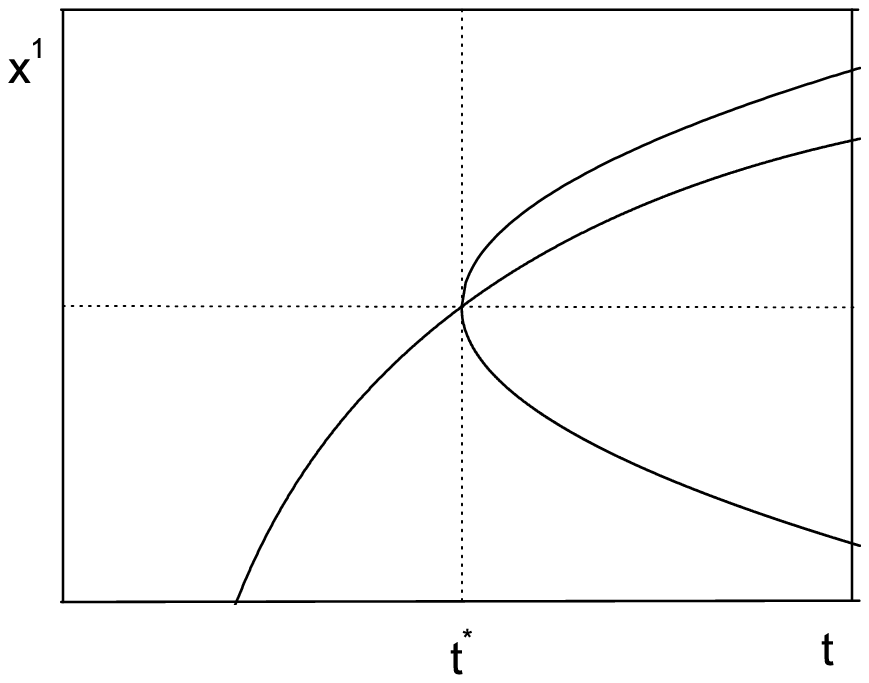,height=5cm,width=5cm}\\
\hspace{2.8cm} (a) & \hspace{2.8cm} (b)
\end{tabular}
\end{center}
\caption{ (a) Three vortices merge into one at the bifurcation
point. (b) One vortices split into three vortices at the bifurcation
point.} \label{fig3}
\end{figure}

\begin{figure}[h]
\begin{center}
\begin{tabular}{p{5cm}p{5cm}}
\psfig{figure=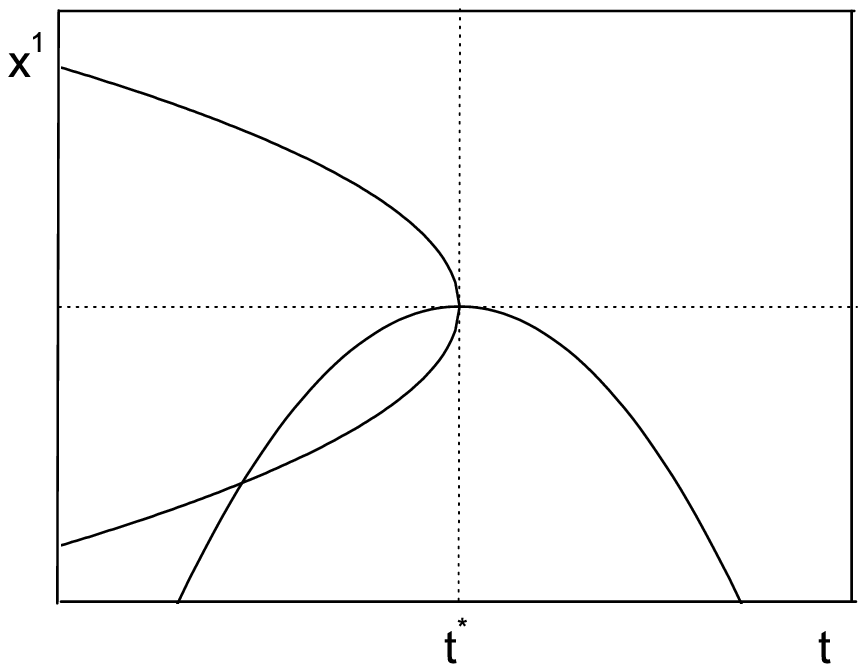,height=5cm,width=5cm} &
\psfig{figure=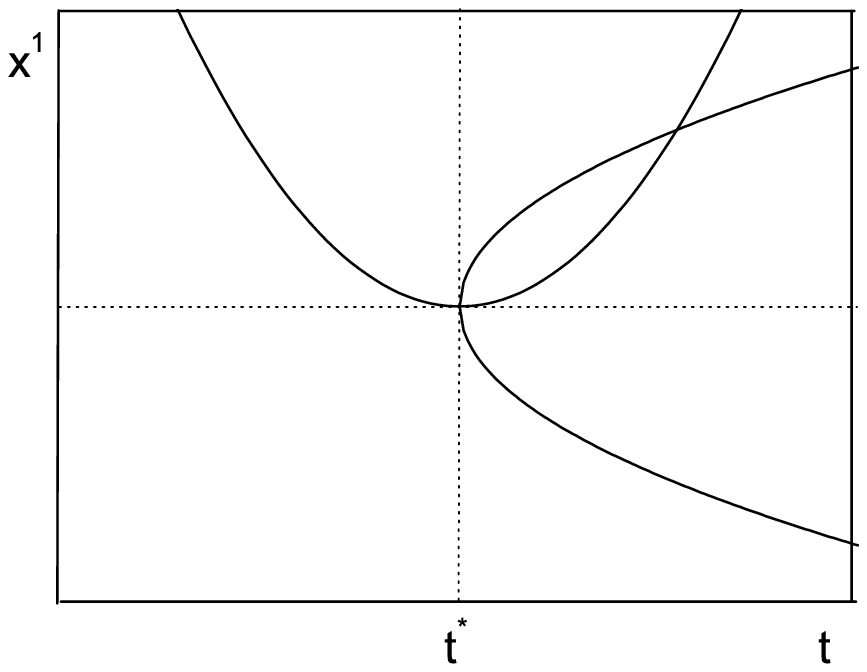,height=5cm,width=5cm}\\
\hspace{2.8cm} (a) & \hspace{2.8cm} (b)
\end{tabular}
\end{center}
\caption{ (a) Three vortices merge into one at the bifurcation
point. (b) One vortices split into three vortices at the bifurcation
point.} \label{fig4}
\end{figure}

Case 4 (A=C=0). Eqs. (\ref{Taylor}) and (\ref{Taylor2}) give
respectively
\begin{equation}
\frac{dx^1}{dt}=0,\ \ \frac{dt}{dx^1}=0.
\end{equation}
This case is obvious, see FIG. \ref{fig4}

These cases reveal the evolution of the self-dual vortices in the
FQH system. Beside two vortices encounter and then depart at the
bifurcation point along different branch curves, it also includes
splitting and merging of vortices. When a multicharged vortex moves
through the bifurcation point, it may split into several vortices
along different branch curves. On the contrary, several vortices can
merge into one vortex at the bifurcation point.

The identical conversation of the topological charges shows the sum
of the topological charges of these final vortices must be equal to
that of the original vortices at the bifurcation point, i.e.,
\begin{equation}
\sum_i\beta_{k_i}\eta_{k_i}=\sum_f\beta_{k_f}\eta_{k_f}.
\end{equation}

Furthermore, from the above discussion, we can see that the
generation, annihilation, and bifurcation of vortices are not
gradually changed, but suddenly changed at the critical points.

\section{conclusion}\label{conclusion}

In conclusion, using the $\phi$-mapping topological current
theory, the topological properties of self-dual vortices in the
FQH system are studied. In Sec. \ref{section1}, we obtain an exact
Bogomol'nyi self-dual equation with topological term, which is
ignored in traditional self-dual equation. It is revealed that
there are self-dual vortices in the FQH system and the topological
charges are determined by Hopf indices and Brouwer degrees. In
Sec. \ref{section2}, we point out that the self-dual vortices
generate or annihilate at the limit points and encounter, split or
merge at the bifurcation. It is shown that the topological charges
of the vortices are preserved in the branch processes during the
evolution of these self-dual vortices.

At last, it should be pointed out that in the present paper we treat
the vortices as mathematical lines, i.e., the width of a vortex line
is zero. This description is obtained in the approximation that the
curvature radius of a vortex line is much larger than the width of
the vortex line \cite{H.B.NielsenNPB1973}.

\section*{ACKNOWLEDGEMENTS}

It is a great pleasure to thank B. H. Gong for numerous fruitful
discussions. This work was supported by the National Natural Science
Foundation of China under Grant No. 10475034.

\end{document}